\documentclass[twocolumn,superscriptaddress,aps,pre]{revtex4-1}

\usepackage{graphicx}
\usepackage{amssymb}
\usepackage{amsmath,amsfonts,epsfig}
\usepackage[english]{babel}
\usepackage{subfig}

\begin{document}

\title{Stabilizing the Hexagonal Close Packed Structure of Hard Spheres with Polymers : Phase diagram, Structure, and Dynamics}

\author{John R Edison}
\affiliation{Soft Condensed Matter, Utrecht University, Princetonplein 5, 3584 CC, Utrecht, The Netherlands}
\affiliation{Present address: Molecular Foundry, Lawrence Berkeley National Laboratory, 1 Cyclotron Road, Berkeley, CA 94720, USA}
\author{Tonnishtha Dasgupta} 
\affiliation{Soft Condensed Matter, Utrecht University, Princetonplein 5, 3584 CC, Utrecht, The Netherlands}

\author{Marjolein Dijkstra}
\affiliation{Soft Condensed Matter, Utrecht University, Princetonplein 5, 3584 CC, Utrecht, The Netherlands}

\email{m.dijkstra@uu.nl}

\begin{abstract}
We study the phase behaviour of a binary mixture of colloidal hard spheres and freely-jointed chains of beads using Monte Carlo simulations. Recently Panagiotopoulos and coworkers predicted [Nat. Commun. 5, 4472 (2014)] that the hexagonal close packed (HCP) structure of hard spheres can be stabilized  in such a mixture due to the interplay between polymer and the void structure in the crystal phase. Their predictions were based on estimates of the free-energy penalty for adding a single hard polymer chain in the HCP and the competing face centered cubic (FCC) phase. Here we calculate the phase diagram using free-energy calculations of the full binary mixture and find a broad fluid-solid coexistence region and a metastable gas-liquid coexistence region. For the colloid-monomer size ratio considered in this work,  we find that the HCP phase is only stable in a small window at relatively high polymer reservoir packing fractions, where the coexisting HCP phase is nearly close packed. Additionally we investigate the structure and dynamic behaviour of these mixtures. 
\end{abstract}

\maketitle

\section{Introduction}
  
Colloidal self-assembly is a promising and viable approach to fabricate new designer materials. Recent advancements in colloidal chemistry have enabled the synthesis of a wide variety of building blocks, with immense control over the size, shape and functionality. This has opened the possibilities to fabricate via self-assembly, a vast variety of hierarchical structures. Very often the targeted structure cannot be accessed due to kinetic traps or due to competing crystal morphologies, which differ very little in free energy. It is the latter problem we deal with in this work. A well-known example of this problem is the crystallization of colloidal particles that are hard-sphere-like. It is well-established that the stable crystal structure of hard spheres is the face centered cubic (FCC) structure \cite{bolhuis1997entropy, mau1999stacking}. The hexagonal close packed (HCP) structure, which differs from the FCC structure only by the stacking order of the hexagonal planes of spheres costs an extra free energy of  $O(10^{-3}) k_B T$ per particle at the melting density. This minute free-energy difference between the two competing structures results in crystallization of the hard-sphere fluid into a random hexagonal close packed (rHCP) phase of which the stacking sequence is random \cite{Pusey1989,Zhu1997crystallization} . Binary mixtures of hard spheres with a diameter ratio of $\simeq 0.8$ show a similar behavior, where three different crystalline polymorphs, the so-called binary Laves phases MgCu$_2$, MgZn$_2$, and MgNi$_2$, differ very little in free energy \cite{hynninen2009stability}. The MgCu$_2$ structure can be used to fabricate photonic crystals with a bandgap in the visible region \cite{hynninen2007self} and therefore a strategy to selectively stabilize this structure is of technological importance.

One approach to target a certain polymorph from its competing structures is to tune the interactions between the colloids in a suitable manner. Significant efforts have been undertaken in the last decade to devise strategies to tune colloidal interactions; some examples are the addition of polymer depletants \cite{lekkerkerker2011colloids} and tuning of salt concentration in systems of charged colloids. Recently, Mahynksi {\em et al} \cite{mahynski2014stabilizing} have shown via computer simulations that in a system of colloids which can exhibit more than one competing crystalline polymorph, addition of non-adsorbing polymers with a carefully chosen architecture can stabilize one polymorph over all the others. This observation has served as the motivation for our work in this paper. Using computer simulations, Mahynski \textit{et al.} showed that the free-energy penalty incurred to add a freely-jointed chain of hard beads to the FCC and HCP crystal of hard spheres, which are kept at the same packing fraction, differed significantly beyond a certain chain length of $M$ beads. The polymers exhibited a clear preference to reside in the void space of one crystal polymorph over the other. Mahynksi \textit{et al} attributed the observed preference, to the difference in distribution of void spaces between the two competing (FCC/HCP) polymorphs in the hard-sphere system. 

While at a fixed colloid packing fraction both HCP and FCC structures possess the same amount of total void space, their distribution in space or connectivity, which places constraints on the possible conformations adopted by the polymer, is very different. Note that by void space we do not refer to vacancies on the crystal lattice, but just the free volume available to the polymer, which  is as large as $26\%$  in a close-packed hard-sphere crystal since the maximum colloid packing fraction is $\eta_{cp}\simeq 0.74$.  The shape of the voids in a hard-sphere crystal can be described by polyhedra. In both the FCC and the HCP structure the void space consists of one octahedral void (OV) and two tetrahedral voids (TV) per particle. The OV is about 6 times larger in volume  than the TV. Furthermore, for both the FCC and HCP structure, the void arrangement is the same in the plane of the hexagonal layers, i.e., OVs only share faces with TVs. However, the crucial difference between FCC and HCP lies in the connectivity of the OVs and TVs in the direction perpendicular to the hexagonal layers. In the FCC lattice, each unit cell has a octahedral void (OV) in the middle capped by smaller tetrahedral voids (TV), and hence the larger OVs in the FCC structure are all isolated from each other. On the other hand  in the HCP structure, the OVs are stacked on top of each other and share faces, whereas TVs are likewise stacked and share either a face or a vertex in the direction perpendicular to the hexagonal planes.

Therefore a hard bead chain which requires more free space than provided by a single OV void, prefers the HCP structure where neighboring OV voids are easily accessible, as they are merely stacked on top of each other. In the FCC structure the same polymer chain has to incur a higher free-energy penalty as it has to find its way through a much narrower TV. Hence Mahynski \textit{et al} predicted that a system of colloids and polymer depletants should display a stable HCP crystal phase under certain conditions. In this work we affirm their predictions by computing the phase diagram of a binary mixture of hard spheres and  freely-jointed chains of hard beads as depletants.

This paper is organized as follows. In section \ref{model} we describe our model of a colloid-polymer mixture and the techniques we use to calculate the phase diagram. In section \ref{results} we present our results on the phase behavior and structure of the model. Finally we discuss the relevance of our studies to experimental studies on colloid-polymer mixtures and summarize our conclusions in section \ref{conclusions}. 

\section{Model and Methods}
\label{model}

We consider a binary mixture consisting of hard spheres with size $\sigma_c$ and freely-jointed hard bead chains. The chains, which model linear homopolymers, are composed of $M$ monomer beads of size $\sigma_m$. The bond length between bonded monomers is constrained to a distance of $\lambda \sigma_m$. All non-bonded interactions of the system are assumed to be hard-sphere-like and set by the no-overlap condition.
\begin{equation}
U(r_{ij}) =
\begin{cases}
\infty \quad r_{ij} < \sigma_{ij}\\
0 \quad \textrm{otherwise}
\end{cases}
\label{eq_non_bonded}
\end{equation}
Here $r_{ij}$ and $\sigma_{ij}$ denote the radial center-of-mass distance and mean diameter between a pair of spheres or beads, $i$ and $j$. The bonded interactions between the beads read
\begin{equation}
U_{bond}(r_{ij}) =
\begin{cases}
\infty \quad r_{ij} < \sigma_{m} \\
0 \quad \sigma_{m} \leq r_{ij} \leq \lambda \sigma_{m} \\
\infty \quad r_{ij} > \lambda \sigma_{m} \\
\end{cases}
\label{eq_bonded}
\end{equation}

In this work, we fix the diameter ratio of the beads and colloids to $ q = \sigma_m / \sigma_c = 1/7 \simeq 0.143$, and set $\lambda = 1.1$ and the chain length to $M=14$ beads. Our methodology to compute the phase diagram directly follows the work of Dijkstra and Evans \cite{dijkstra1999direct,dijkstra1999phase}. To compute the phase diagram it is convenient to study the system in the fixed $\{N_c, z_p, V, T\}$ ensemble. Here $N_c$ denotes the number of colloids, $z_p=\exp{\lbrack\beta \mu_p\rbrack}/\Lambda_p^3$ is the fugacity of the hard bead chains, $\mu_p$ the chemical potential of the bead chains, $T$ is the temperature, $\beta=1/k_BT$, and $\Lambda_p$ the thermal wavelength of the polymer, and $V$ is the volume. The exact free energy of the system in this ensemble $F(N_c, z_p, V, T)$, is given by the following identity. 
\begin{widetext}
\begin{eqnarray}
\beta F (N_c, z_p, V)& = &\beta F(N_c, z_p = 0, V) + \int_{0}^{z_{p}}\left ( \frac{\partial \beta F(N_c, z_p^{'}, V)}{\partial z_{p}^{'}} \right ) d z_{p}^{'} = \beta F (N_c, z_p = 0, V) -  \int_{0}^{z_p} \frac{\langle N_p \rangle_{z_{p}^{'}}}{z_p^{'}} d z_{p}^{'}
\label{eq_FE}
\end{eqnarray}
\end{widetext}

The first term of the right hand side is just the Helmholtz free energy of a pure system of $N_c$ colloidal hard spheres  in a volume $V$. We use the Carnahan-Starling free-energy expression for the fluid phase, while for the solid phase we use the Frenkel-Ladd method to obtain accurate estimates of the free energies of FCC and HCP crystal phases of system size $N_c=108$ spheres. The second term is the excess contribution that arises due to the presence of hard bead chains in the system, where $\langle N_p \rangle$ is the average number of polymers present in a system with $N_c$ colloids in equilibrium with a polymer reservoir fixed at fugacity $z_p$. The ``adsorption'' of polymer chains onto a system of $N_c$ colloids in a volume $V$ can be measured directly in a Monte Carlo simulation, where the polymer is treated grand canonically. Subsequently, we estimate the densities of the coexisting phases at a given $z_p$ by performing common tangent construction on the resulting $\beta F (\eta_c,z_p) / V$ vs $\eta_c$ data. More details can be found in the Appendix of Ref. \cite{dijkstra1999phase}. 

\subsection{Configurational Bias Monte Carlo Method}
\label{cbmc}

As stated in the previous section, to compute the Helmholtz free energy $\beta F(N_c,z_p,V)$, we measure  the ``polymer adsorption'' $\langle N_p \rangle$ onto a system of $N_c$ colloids in a volume $V$  in a Monte Carlo simulation at fixed  polymer fugacity $z_p$. In simulations of chain molecules the biggest bottleneck lies in generating trial configurations that are likely to be accepted. A vast majority of the trial configurations result in overlaps with other chain molecules or colloids. One strategy to overcome this issue is to bias the generation of trial configurations. The bias can then be accounted for by modifying the Monte Carlo acceptance factors suitably. The configurational bias Monte Carlo (CBMC) method uses the Rosenbluth scheme to generate trial configurations.  In this method, we grow the polymer chain bead by bead. In order to add a bead, we generate first a fixed number of trial positions, say k, and calculate the associated Boltzmann weight. We choose a certain trial position among the k options, with a probability proportional to its Boltzmann weight. In this way, we generate trial configurations that are more likely to be accepted. More details on the implementation of this method can be in found in the following references \cite{Smit1995, Frenkel2001}. We employ CBMC in our simulations to insert/delete/translate the polymer chains. A single sweep in our simulations consists of the following steps i) translation moves of all particles in the system ii)  2000 configurational bias moves which are equally split between polymer translation, insertion and deletion.  

The measurement of $\langle N_p \rangle$  requires a simultaneous canonical averaging of the colloid (big species) configurations. At the chosen colloid-monomer size ratio $q=1/7$ we find that the colloids sample the configuration space efficiently, with just single particle moves. At a few state points we have employed the rejection-free event-chain Monte Carlo \cite{Bernard2009} algorithm,  which helps the colloids to sample the configuration space more efficiently. The adsorption isotherms computed with and without the event chain moves are indiscernible. This indicates that the rate limiting steps for equilibration are the polymer addition/deletion moves. 

\subsection{Transition Matrix Monte Carlo method}
\label{tmmc}

To obtain an accurate estimate of the adsorption of polymer onto a system of $N_c$ colloids in a volume $V$ as a function of fugacity   $z_p$, we employ the Transition Matrix Monte Carlo (TMMC) method, developed by Fitzgerald and coworkers \cite{Errington2003,Fitzgerald1999}.  In a typical simulation at fixed $\{N_c, z_p, V, T\}$, the number of polymers $N_p$ fluctuates around an  equilibrium value. Using TMMC, we measure the probability of observing a certain number of polymers in the system $P(N_p)$.  The technique involves performing the same simulation as mentioned above in section \ref{cbmc} together with a few bookkeeping steps.

After every insertion $(N_p \rightarrow N_p+1)$ and deletion$(N_p \rightarrow N_p-1)$ in a configurational bias Monte Carlo move for a polymer chain, we update a collection matrix $C$ in the following fashion.
\begin{widetext}
\begin{align}
\label{eq02}
C (N_p \rightarrow N_p + \delta N_p) &= C(N_p \rightarrow N_p+\delta N_p) + a(N_p \rightarrow N_p + \delta N_p)  \nonumber \\
C (N_p \rightarrow N_p) &= C(N_p \rightarrow N_p) + 1 - a(N_p \rightarrow N_p+\delta N_p)  
\end{align}
\end{widetext}
Here $a$ is the configurational bias acceptance probability of the proposed move. Note that the collection matrix is updated regardless of the move being accepted or rejected. The dimension of the collection matrix is $\{3 , N_p^{max} + 1\}$, where $N_p^{max}$, is a chosen upper limit of the range we wish to sample. Since we attempt to add or remove a single chain in each configurational bias Monte Carlo move,  $\delta N_p$ is $\pm 1$. Periodically during the simulation we use the information obtained via bookkeeping to compute a biasing function $\Phi(N_p)$. To compute $\Phi(N_p)$, we estimate transition probabilities $W(N_p \rightarrow N_P^{'})$ using the data in the collection matrix $C$. 

\begin{equation}
\label{eq03}
W(N_p \rightarrow N_P^{'}) = \frac{C(N_p \rightarrow N_P^{'})}{\sum_{N_p^{''}} C(N_p \rightarrow N_p^{''})}
\end{equation}

The summation runs over the three possible states of $N_p^{'}$, which are $ N_p + 1, N_p - 1$ and  $N_p$. The Monte Carlo detailed balance expression is then employed to estimate the probability distribution $P(N_p)$.

\begin{equation}
\label{eq04}
P(N_p) W(N_p \rightarrow N_P^{'}) = P(N_p^{'}) W(N_p^{'} \rightarrow N_P)
\end{equation}

The biasing function is given by, $\Phi(N_p) = -\ln (P(N_p))$ and the proposed polymer addition or removal moves are then accepted  or rejected  based on a biased acceptance criterion, $ \min \left\{ 1, a \exp(\Phi(N_p + \delta N_p) - \Phi(N_p)) \right\}$. Note that the collection matrix $C$ is always updated with the unbiased acceptance probabilities. The simulation continues until the biasing function (probability distribution) converges within a preset tolerance. A more detailed description of this method can be found in the work of Errington \cite{errington2003direct,Errington2003}. The output from a TMMC simulation is the probability of observing a certain  number of polymer $N_p$ in the system at a fixed fugacity $z_p$. We use this information to estimate the probability distribution at any other value of fugacity $z_p^{'}$ using the histogram re-weighting technique \cite{Panagiotopoulos2000reweight}. The average number of polymers $<N_p>$ at a fixed fugacity $z_p{'}$, is the first moment of the reweighted distribution $P(N_p; z_p^{'})$. 


\section{Results}
\label{results}

\subsection{Polymer adsorption as a function of colloid packing fraction}

\begin{figure}
\centering
\includegraphics[width=3.0in]{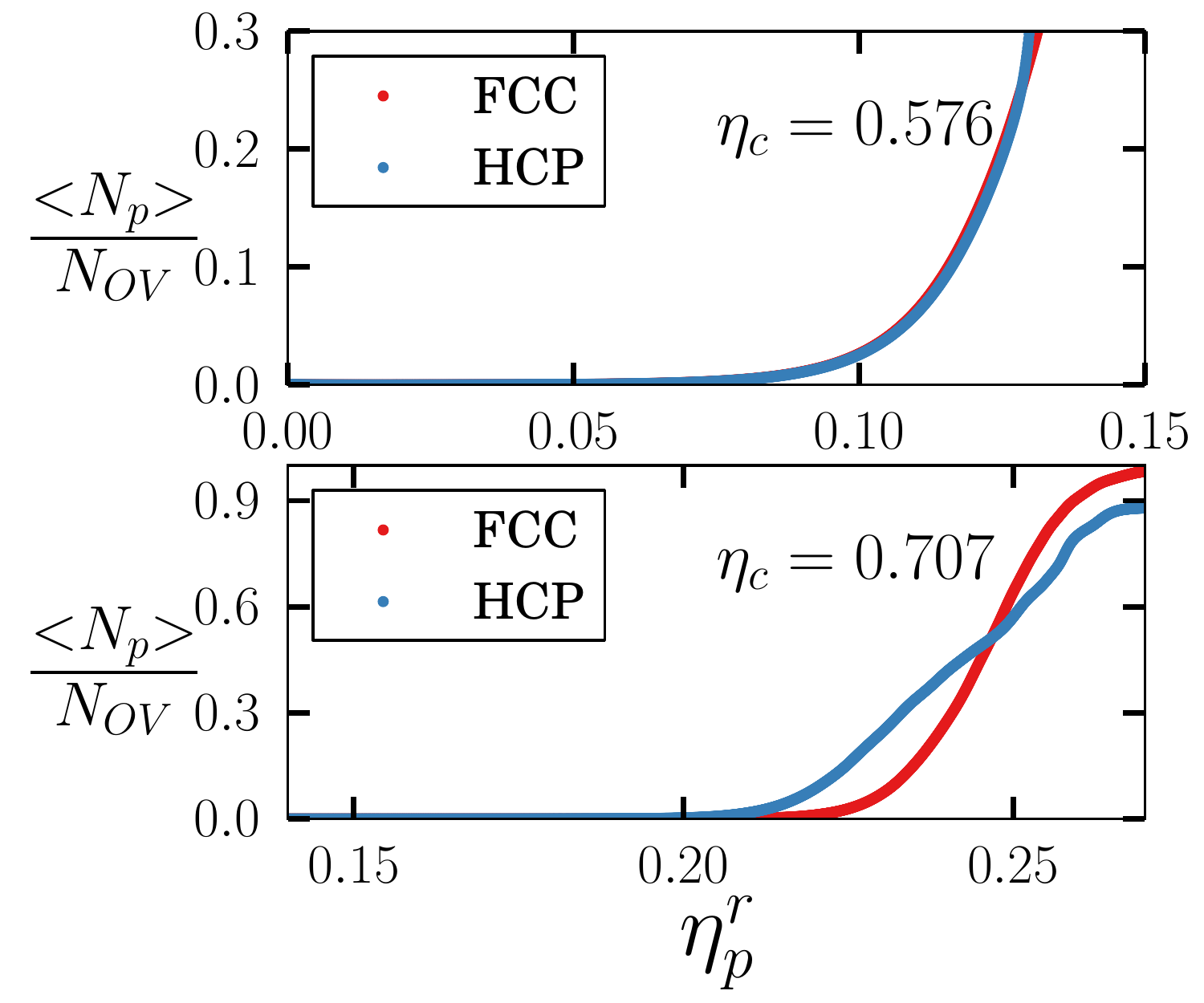}
\caption{\label{fig01}The average number of adsorbed polymer per octahedral void $<N_p>/N_{OV}$ in the HCP (blue) and FCC (red) polymorphs as a function of the polymer reservoir packing fraction $\eta_p^r$ at two different colloid packing fractions  $\eta_c = 0.576$ (top) and $\eta_c = 0.707$ (bottom).}
\end{figure}

As mentioned in Section \ref{model} we fix the number of beads of the polymer chains at $M=14$, and the ratio of the size of the monomer and colloid  at $q = \sigma_m/\sigma_c= 1/7\simeq 0.143$. We choose these specific values  of $M$ and $q$, because previous studies showed that the free-energy penalty to insert such a  polymer  greatly favours the HCP structure  at sufficiently high colloid packing fraction $\eta_c$  \cite{mahynski2014stabilizing}. We treat the polymer chains in the grand canonical ensemble, i.e., we fix the polymer fugacity $z_p$. We also fix the colloid packing fraction $\eta_c$. In simulations where the colloids are crystalline we fix the number of colloids as $N_c = 108$, and choose box-dimensions accordingly. For the fluid phase we fix the linear dimension of the cubic box as $L=30 \sigma_m$ and choose $N_c$. We measure the average number of polymer chains $\langle N_p \rangle$ per octahedral void as a function of polymer fugacity $z_p$ for varying colloid packing fractions $\eta_c$. 

In Fig. \ref{fig01} we show the adsorption isotherms or the number of polymers adsorbed per octahedral void $\langle N_p \rangle/N_{OV}$ as a function of the reservoir polymer packing fraction  $\eta_p^r$ at fixed colloid packing fraction $\eta_c=0.576$ and 0.707. Here $\eta_p^r$ is the packing fraction of a reservoir of pure polymers at fugacity $z_p$, $ \langle N_p \rangle$ is the number of polymers adsorbed onto the system, and $N_{OV}$ is the number of octahedral voids in the system. . We determine this conversion by performing Monte Carlo simulations on a pure system of bead chains at fixed polymer fugacity $z_p$ and measuring the corresponding polymer packing fraction.  At colloid packing fraction $\eta_c=0.576$ (Fig. \ref{fig01} top) we observe no significant difference in the adsorption isotherms of the FCC and HCP structures. The different void distributions in the FCC and HCP structure do not result in a noticeable difference in the polymer adsorption. At this value of $\eta_c$, there appears to be plenty of space available for the polymers, and the colloids do not appear to constrain the chain configurations.  It is worth noting that at $\eta_c \leq 0.576$, phase separation was observed in the simulation box for values of $\eta_p^r$ much higher than shown in Fig. \ref{fig01}a. Hence,  there is an upper bound in $\eta_p^r$ for estimating the free energy $F(N_c,z_p,V,T)$ at fixed $\eta_c$.

At colloid packing fraction $\eta_c = 0.707$ (Fig. \ref{fig01} bottom) the adsorption isotherms display two interesting features: i) At $\eta_p^r \simeq 0.225$, the polymer adsorption in the HCP structure is much higher than for the FCC structure. This can be explained by the fact that  the size of the polymer as given by the radius of gyration exceeds the size of the octahedral hole, forcing the polymer to venture to neighbouring voids. Hence, the polymer is forced to explore the smaller tetrahedral voids in the case of the  FCC structure, thereby incurring a significant free-energy penalty, while in the HCP structure, the chains reach into the larger octahedral voids. This is the reason why the polymer adsorption is significantly higher in the HCP than in the  FCC phase at low $\eta_p^r$ . ii) Upon further increasing $\eta_p^r$, the polymer adsorption in the FCC structure becomes larger than in the HCP structure. This is because in the HCP structure 
more than one chain occupies an octahedral void. Therefore as $\eta_p^r$ increases, FCC wins over again, as its OV's are left mostly unoccupied, while in the HCP structure the polymers are forced to share the octahedral void with another chain . 

As mentioned earlier in section \ref{model} the average number of polymer chains $\langle N_p \rangle / N_{OV}$ per octahedral void allows us to calculate  the Helmholtz free energy $F(N_c,z_p,V,T)$. Our adsorption isotherms together with Eq. \ref{eq_FE} clearly indicate that there is a region in the phase diagram of this binary mixture of hard spheres and hard polymer chains, where the stable crystal phase is the HCP phase --- this is what we intend to establish with free-energy calculations in the next subsection. 

\subsection{Phase Behaviour}	

\begin{figure*}
\centering
\subfloat[]{\includegraphics[width=3.5in]{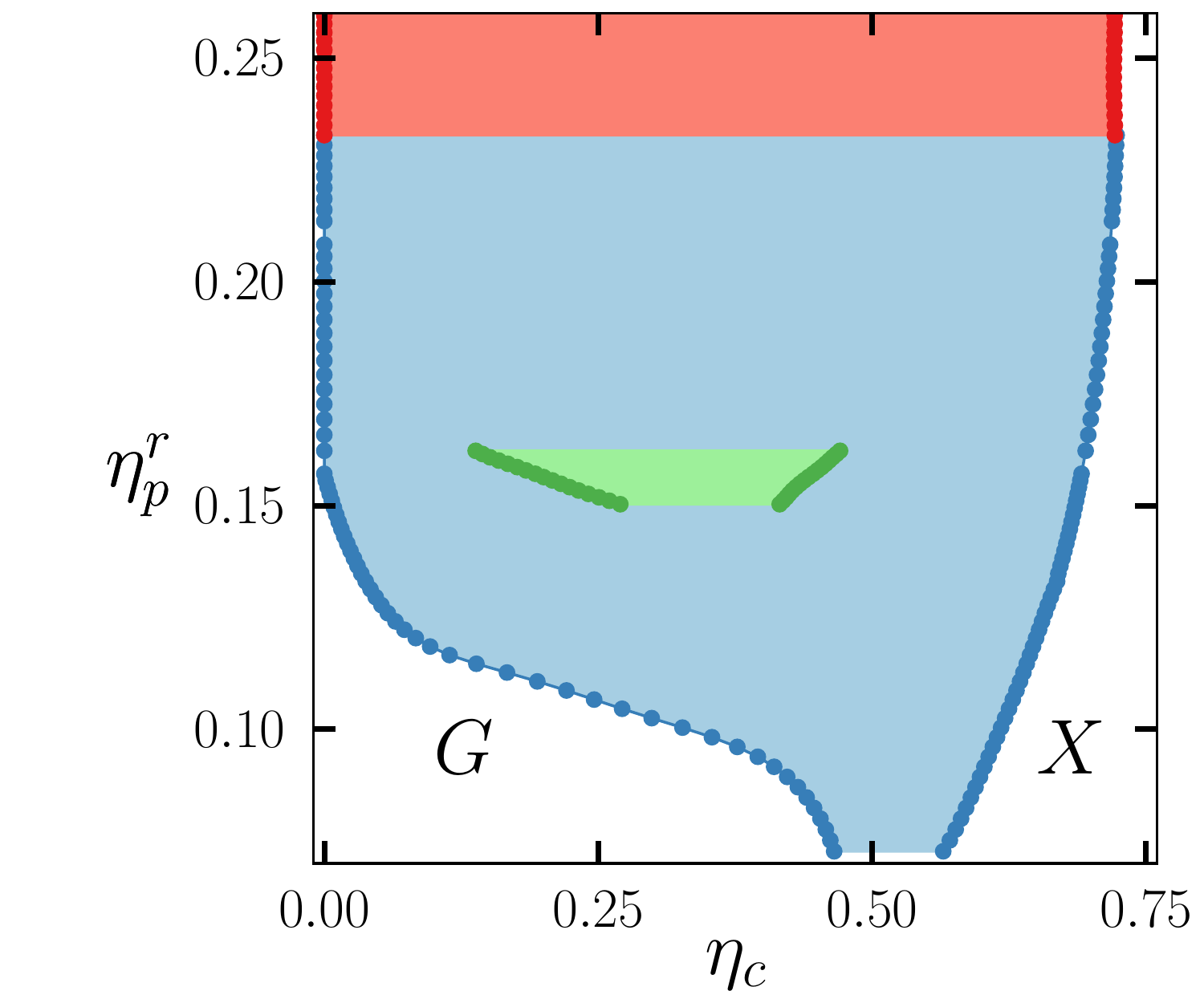}\label{fig02a}}
\subfloat[]{\includegraphics[width=3.5in]{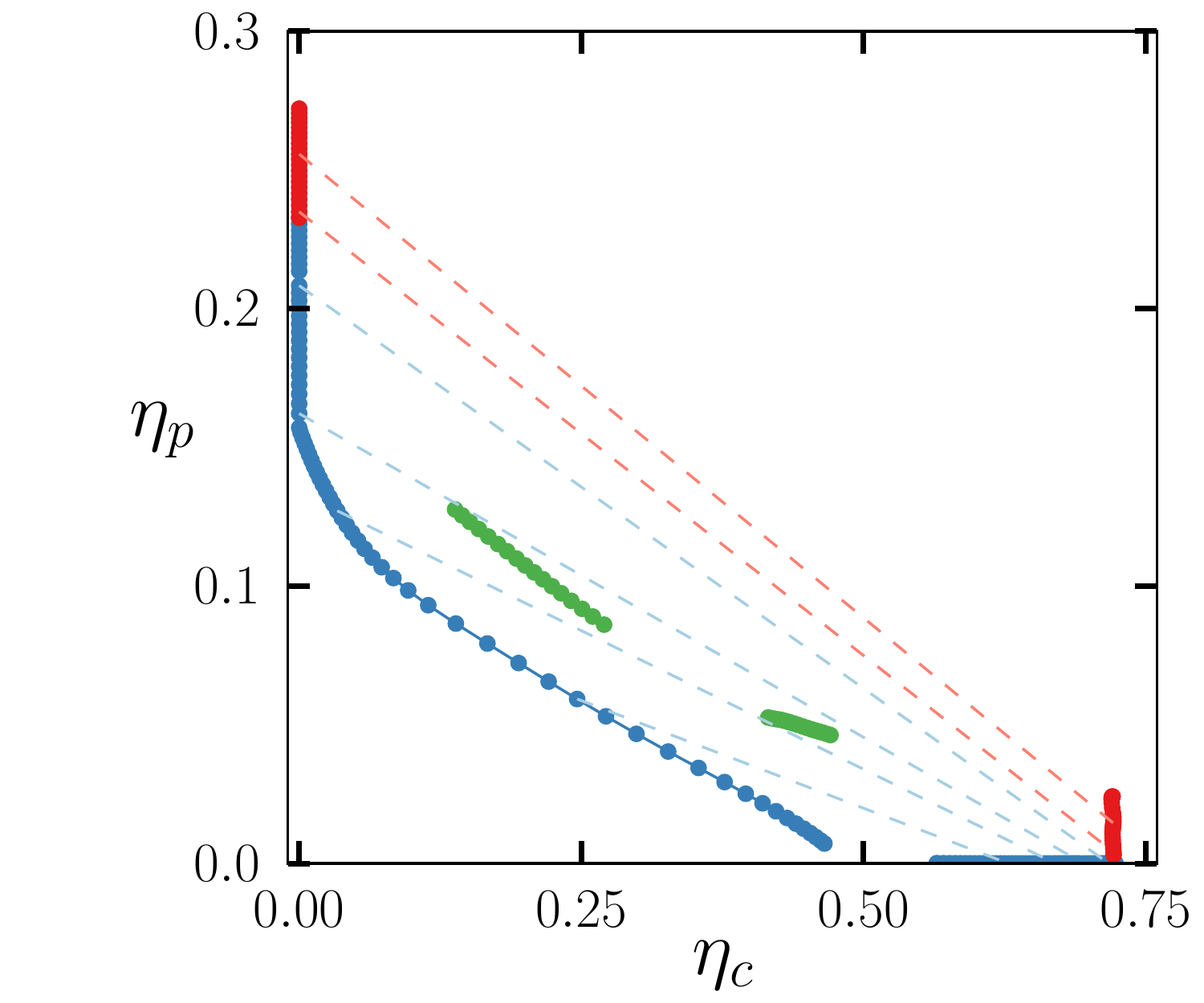}\label{fig02b}} \\
\subfloat[]{\includegraphics[width=2.25in]{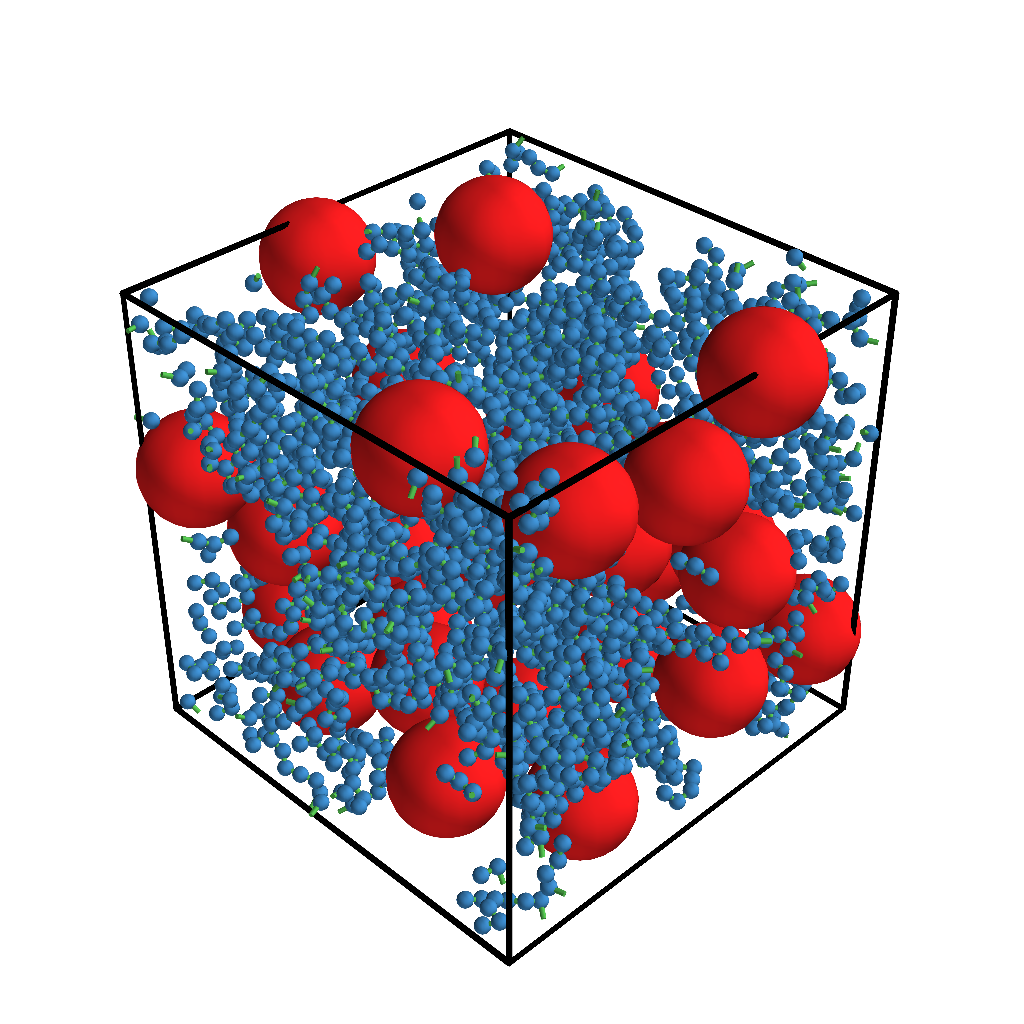}\label{fig02c}}
\subfloat[]{\includegraphics[width=2.25in]{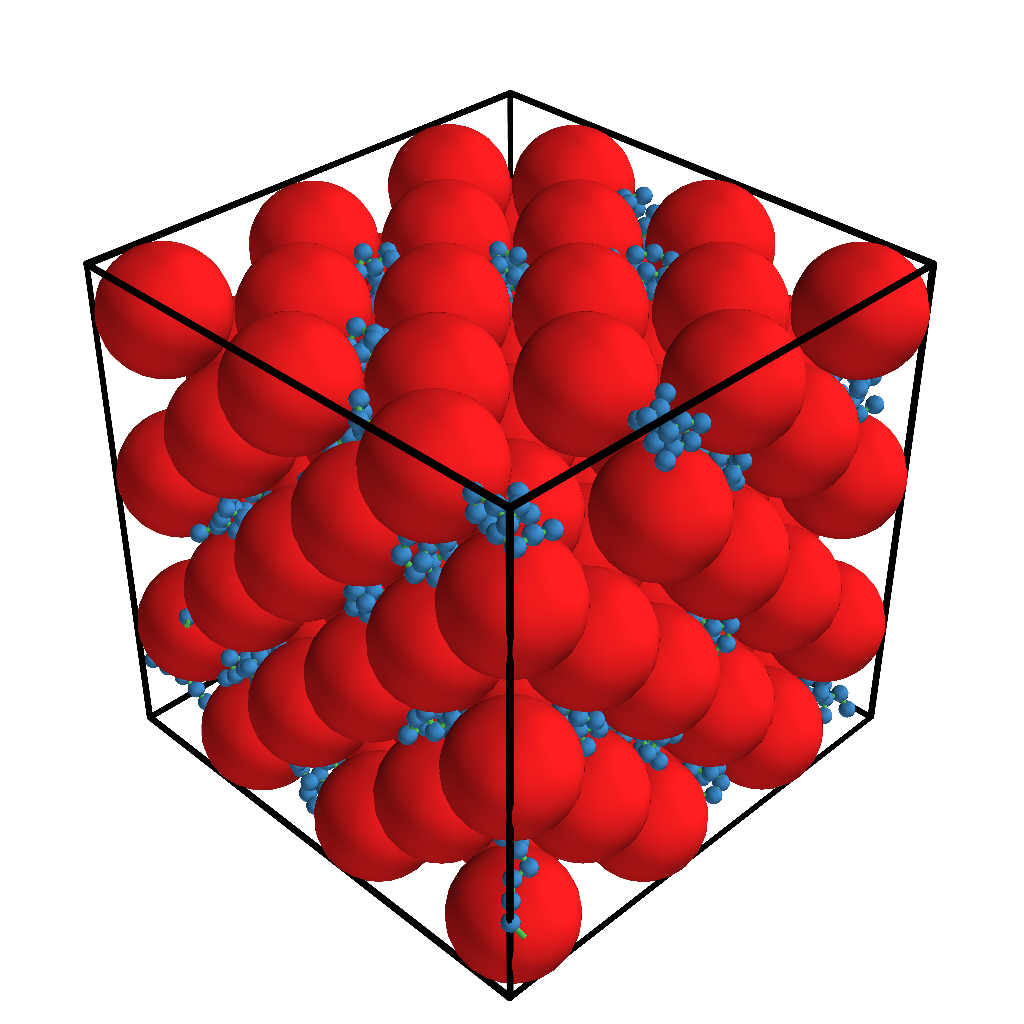}\label{fig02d}}
\subfloat[]{\includegraphics[width=2.25in]{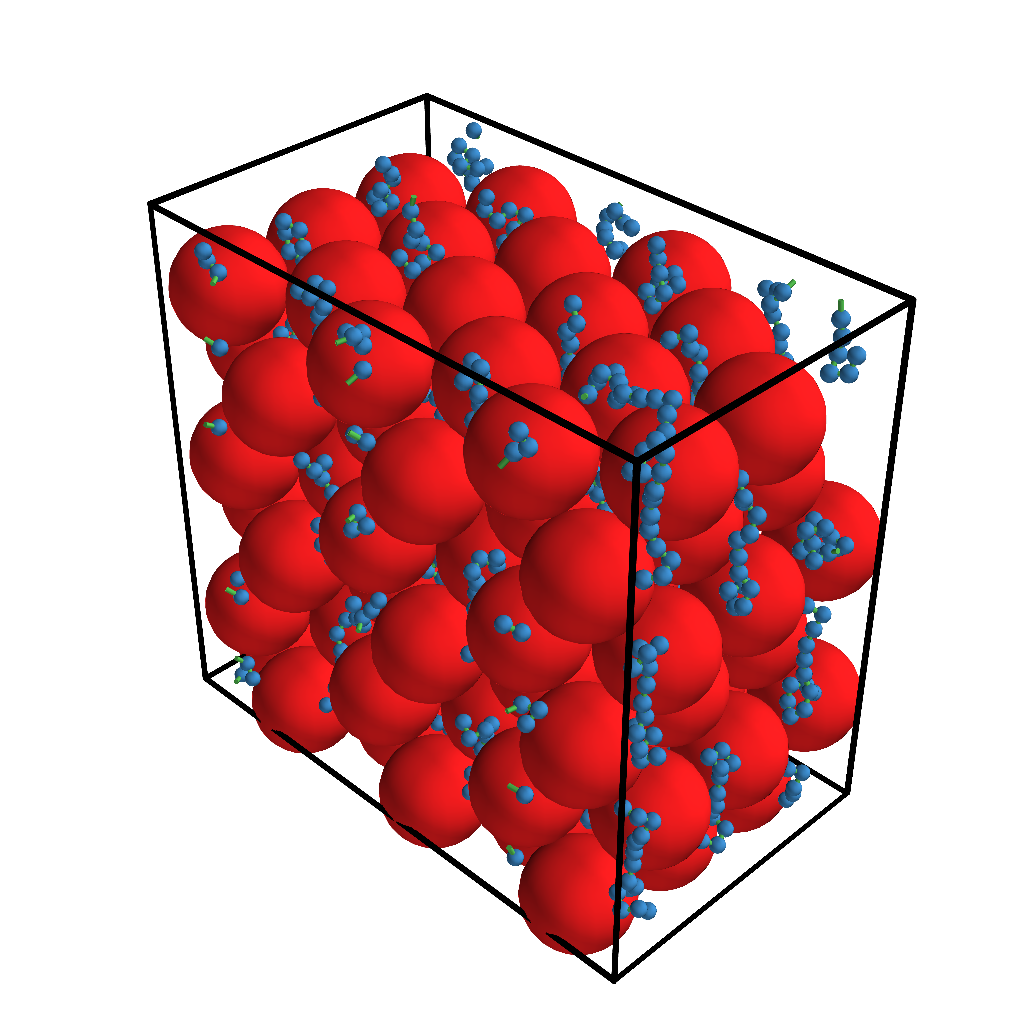}\label{fig02e}}
\caption{\label{fig02} a. Phase diagram of a binary mixture of colloidal hard spheres and freely-jointed polymer chains of $M=14$ beads and  a size ratio $q=1/7$  in the polymer reservoir packing fraction  $\eta_p^r$ vs colloid packing fraction $\eta_c$ representation. b. Phase diagram in the $\eta_p$ vs $\eta_c$ representation. The tie-lines connect the coexisting phases. Typical confiigurations of the colloid-polymer mixture representative of c) fluid phase $N_c=30$, $\eta_p^r = 0.096$ d) crystal phase with FCC ordering  $N_c=108$, $\eta_p^r = 0.254$ and e) crystal phase with HCP ordering $N_c=108$, $\eta_p^r = 0.254 $}
\end{figure*}
 
 Using the data as obtained from the adsorption isotherms, $\langle N_p \rangle$ as a function of $\eta_p^r$ or $z_p$, we calculate $F(N_c,z_p,V)$ using Eq. \ref{eq_FE} for the colloidal fluid phase, and the HCP/FCC crystal phases. We employ common tangent constructions at fixed $z_p$ to determine the phase boundaries. In Fig. \ref{fig02}a we plot the phase diagram of the colloid-polymer mixture in the colloid packing fraction $\eta_c$ vs polymer reservoir packing fraction $\eta_p^r$ representation. In this representation tie lines that connect the two coexisting phases are horizontal. At $\eta_p^r=0$, the coexisting densities are simply given by the fluid-solid transition of pure hard spheres. Upon increasing $\eta_p^r$ ($\mu_p$), an enormous  broadening of the fluid-solid  transition is observed. Moreover, we also find a metastable gas-liquid (G-L) phase coexistence. This metastable G-L phase coexistence terminates in a critical point, but we have not been able to accurately locate the critical point due to its metastability. 

It is worthwhile to compare  the phase diagram as shown in Fig. \ref{fig02}a with the phase diagram of a binary hard-sphere mixture with the same size ratio $q = 1/7$. In  Ref. \cite{dijkstra1999phase}, phase diagrams  are reported for size ratio $q=1/10$ and $q=1/5$. We observe that the topology of the phase diagram shows a striking resemblance with the phase diagrams for binary hard-sphere mixtures, i.e., both phase diagrams show a metastable gas-liquid transition and an enormous widening of the fluid-solid transition at a remarkably similar polymer reservoir packing fraction, see Figs. 14a and 14b of Ref. \cite{dijkstra1999phase}. This latter finding is particularly striking as the reservoir packing fraction at which the widening of the fluid-solid transition occurs is similar for both the single-sphere depletant and the polymer depletant, with chain connectivity playing no role. In Fig. \ref{fig02}b we replot the phase diagram in the $\eta_p$ vs $\eta_c$ representation with $\eta_p$ the actual polymer packing fraction in the coexisting phases. We also show tie lines  in Fig. \ref{fig02}b as denoted by the slanted dashed lines that connect the two coexisting phases. 

\begin{figure}
\centering
\includegraphics[width=3in]{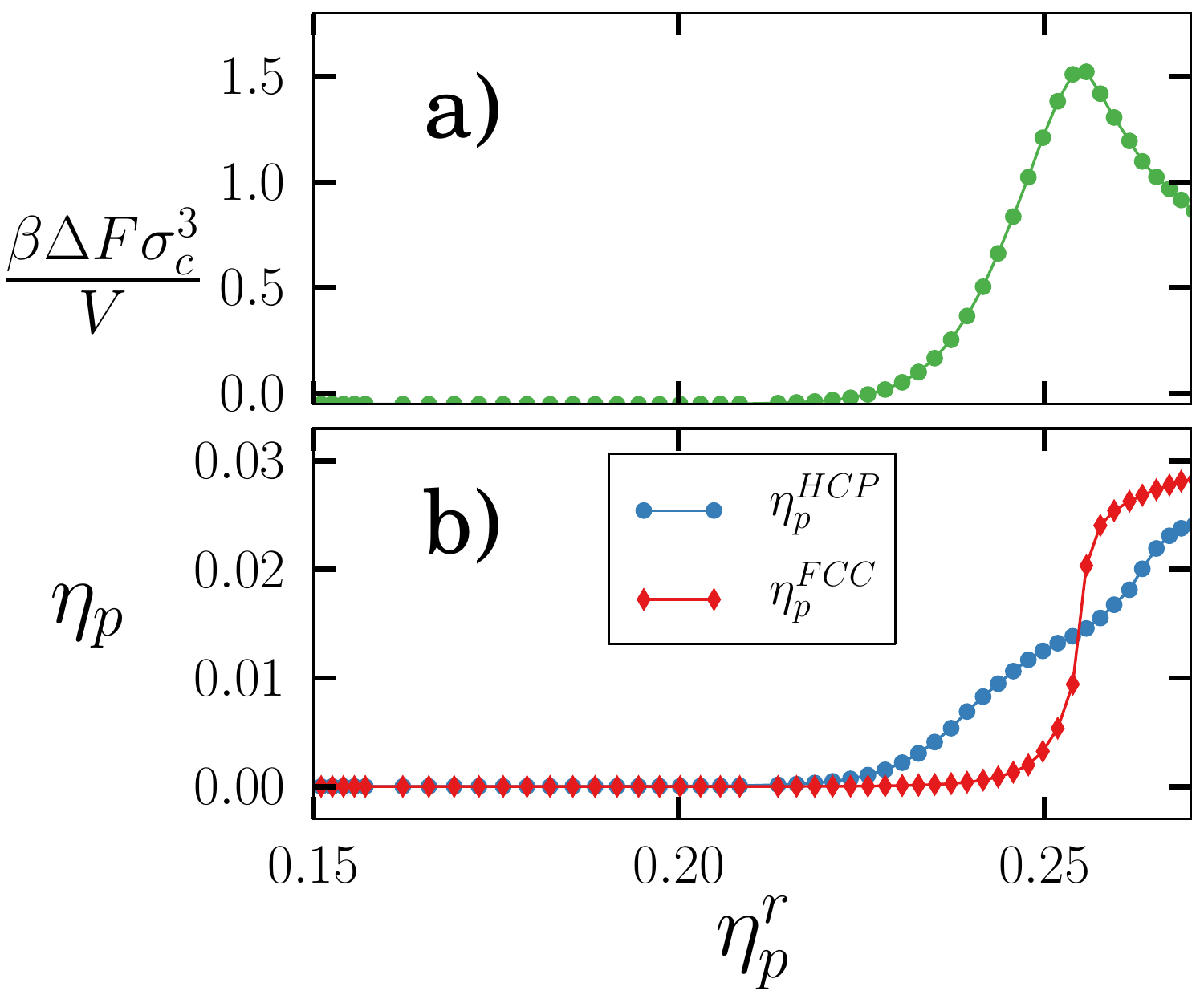} 
\caption{\label{fig03} a) Free-energy difference between the FCC and HCP crystal structure that are in coexistence with the fluid phase,  $\beta (F_{FCC}^{cx} - F_{HCP}^{cx}) \sigma_c^3/ V$, as a function of  the polymer reservoir packing fraction  $\eta_p^r$. b) The polymer packing fraction in the FCC (red symbols) and HCP (blue) crystal phase (blue symbols) that are in coexistence with the fluid phase vs the polymer reservoir packing fraction  $\eta_p^r$.}
\end{figure}

In the limit $\eta_p^r = 0$, i.e., pure hard spheres,  the stable crystal phase is the FCC phase. As can be seen from Fig. \ref{fig02}b and \ref{fig03}b the amount of polymer in the coexisting crystal phase is negligible up to about $\eta_p^r \simeq 0.225$, and therefore the FCC structure is stable for values of $\eta_p^r < 0.225$. Upon increasing $\eta_p^r$ further, the FCC crystal phase becomes metastable with respect to the HCP phase for $\eta_p^r > 0.225$. Hence, a two-phase coexistence region between a fluid and an HCP crystal appears in the phase diagram as denoted by the red region in Figs. \ref{fig02}a and \ref{fig02}b. 

In Figs. \ref{fig02}c-\ref{fig02}e we show typical configurations of the fluid phase, the metastable FCC crystal phase, and the stable HCP crystal phase. It is interesting to note that Fig. \ref{fig02}d shows clearly that the polymer chains in the octahedral voids of the FCC crystal are compact, whereas they are stretched in the HCP phase thereby occupying more than one octahedral void.  

The top panel in Fig. \ref{fig03} shows the difference in free energy between the FCC and HCP crystal phases that are in coexistence with a colloidal fluid phase as a function of the packing fraction of the polymer $\eta_p^r$ in the bulk reservoir. It is clear that for $\eta_{p}^{r} > 0.225$, the HCP structure is the stable phase, with a free-energy difference becoming more than $~1 ~k_BT$ per particle. Upon further increasing $\eta_p^r$, the free-energy difference begins to fall. The difference in free energy between the colloid-polymer FCC and HCP crystal structure is directly related to the difference in polymer adsorption as expected from Eq. \ref{eq_FE}. In Fig. \ref{fig03} (lower panel) we plot the polymer packing fraction in both the HCP (stable/metastable) and FCC (stable/metastable) crystal phase that are in coexistence with the fluid phase. As can be seen from the lower panel in Fig. \ref{fig03}, the polymer adsorption onto the  FCC crystal wins over that of the HCP phase at large $\eta_p^r$. Based on this and the data shown in Fig. \ref{fig01} we expect the FCC crystal structure to become stable  again at very high values of $\eta_p^r$. However, simulations at these state points are infeasible. 

\subsection{Structure and Dynamics}

\begin{figure}
\centering
\includegraphics[width=3.0in]{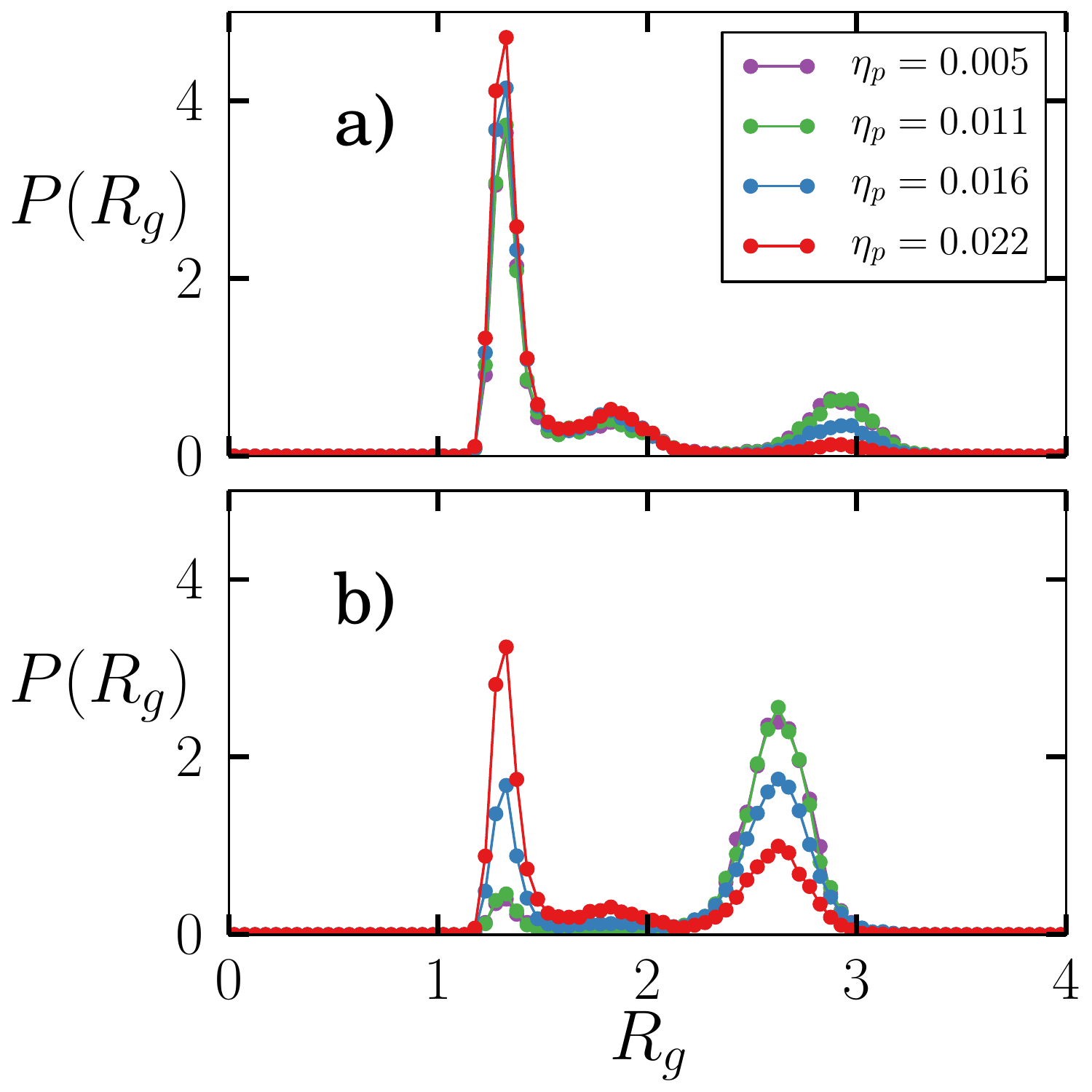}
\caption{\label{fig04} Probability distribution of the radius of gyration $P(R_g)$ of the polymer chains with  length $M=14$ beads for varying polymer packing fraction ($\eta_p$) as indicated in the legend in a ) the FCC structure and b) the HCP structure. The packing fraction of the colloid is fixed at $\eta_c = 0.722$. The radius of gyration $R_g$ is expressed in units of monomer size $\sigma_m$.}
\end{figure}

\begin{figure*}
\centering
\subfloat[]{\includegraphics[width=1.50in]{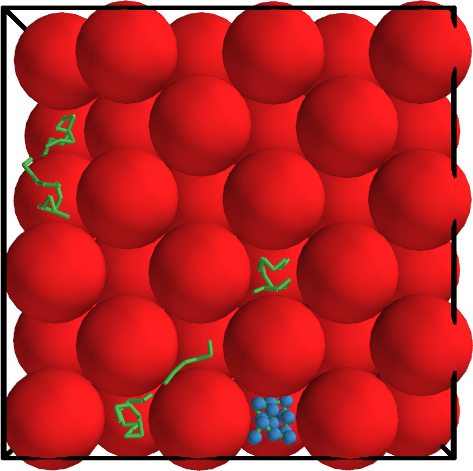}\label{fig05a}} \qquad
\subfloat[]{\includegraphics[width=1.50in]{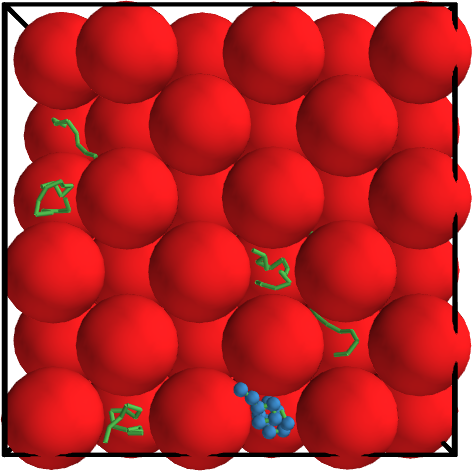}\label{fig05b}} \qquad
\subfloat[]{\includegraphics[width=1.50in]{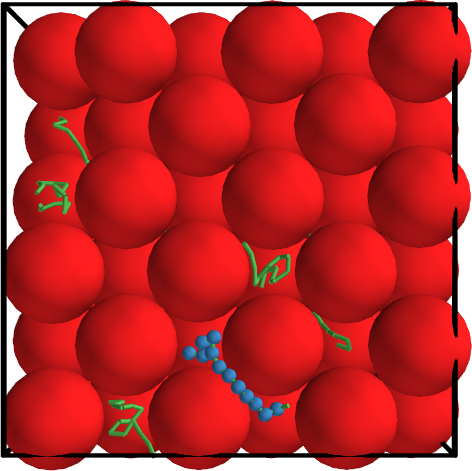}\label{fig05c}} \qquad
\subfloat[]{\includegraphics[width=1.50in]{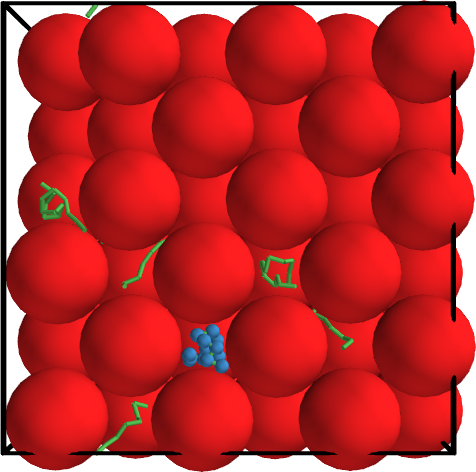}\label{fig05d}}
\caption{\label{fig05} Snapshots from an event driven molecular dynamics simulation which show the activated jump of a polymer chain of length $M=14$ (shown in blue) from one octahedral void to another in the FCC lattice. The packing fraction of the colloid is fixed at $\eta_c = 0.722$.}
\end{figure*}

Next we take a closer look at the structure of the polymers in the FCC and HCP polymorphs at a fixed value of the packing fraction of the colloids $\eta_c = 0.722$. Using event driven molecular dynamics simulations, we investigate the structure and dynamics of the polymers at fixed $\{ N_c, N_p, V, T\}$. Note that at relatively high polymer adsorption the FCC structure is metastable, and should transform into the HCP structure in the long time limit. However the barrier associated with changing the stacking sequence of the hexagonal planes as well as  the fixed shape of the simulation box, allows us to simulate the metastable FCC phase without a spontaneous transformation to the HCP phase. In Fig, \ref{fig04} we plot the probability distribution of the radius of gyration $R_g$ (in units of monomer size $\sigma_m$) of the polymer chains adsorbed onto the HCP and FCC polymorphs as a function of the polymer packing fraction $\eta_p$. From Figs. \ref{fig02}e and \ref{fig04}a, it is evident that the polymers stay collapsed within the octahedral holes in the FCC polymorph. At low to intermediate polymer packing fractions $\eta_p$, the polymers diffuse to the neighbouring octahedral cavity in the FCC crystal phase. This diffusion process occurs via the polymers stretching into a tetrahedral cavity to be able to translocate from one octahedral cavity to another. These jumps are activated as they involve an entropic barrier associated with the entropic penalty the chains incur, as they stretch into a tetrahedral cavity before jumping into the neighbouring octahedral cavity. In Fig. \ref{fig05} we show snapshots from an EDMD simulation which shows a polymer chain (shown in blue) performing an activated jump between two octahedral voids. It might be of interest to investigate the statistics of the translocation times of the hard bead chains between neighbouring cavities, which must be coupled to the vibrations of the colloidal crystal lattice.  

In the HCP polymorph, the polymer chains are more stretched at low polymer packing fraction and as a consequence the polymers occupy more than one octahedral cavity. However, the polymers do not appear to be freely migrating in the columns formed by the octahedral cavities of the HCP structure. The probability distribution of the radius of gyration as shown in figure \ref{fig04} shows a single peak at  $R_g \simeq 2.7$. However upon increasing the polymer adsorption $N_p / N_{OV} \geq 0.5$ or $\eta_p \geq 0.016 $, effects of crowding set in. This is reflected in the probability distribution of the radius of gyration of the chains which becomes bimodal. The entropy of the system appears to be maximized by the cooperative action of the polymer chains. What happens at these loadings is that a few polymers are in the collapsed state, thereby letting the rest of the chains exist in a stretched state. In supplementary movies S1 and S2 we present animated visualizations of the polymers in the FCC and HCP crystal structure as obtained from our EDMD simulations. 

\section{Discussion and Conclusions}
\label{conclusions}

In conclusion, we have investigated the phase behaviour of a binary mixture of colloid hard spheres with a diameter $\sigma_c$ and freely-jointed bead chains consisting of hard spherical beads with a diameter $\sigma_m$. We determined the phase diagram using free-energy calculations for a fixed monomer-colloid size ratio $q=\sigma_m/\sigma_c=1/7$ and a chain length of $M=14$ beads. We implemented the configurational bias Monte Carlo method to speed up equilibration of  the polymer chains and the transition matrix Monte Carlo method to determine accurately the adsorption isotherms of the polymer. The phase diagram displays a broad fluid-solid two-phase coexistence region and a gas-liquid coexistence region, which is metastable with respect to the broad fluid-solid transition. We find that the FCC crystal structure is stable at low polymer reservoir packing fraction $\eta_p^r$, whereas the HCP structure becomes stable for  $\eta_p^r>0.225$.   Our results further suggests that the FCC structure regains stability at very high values of the polymer reservoir packing fraction $\eta_p^r$. We also studied the structure and dynamics of the polymers in the crystal phase. In the FCC structure, the polymers stay collapsed in an octahedral cavity performing activated jumps to neighboring cavities. It is worth mentioning that a similar behavior was observed for an interstitial solid solution in a binary hard-sphere mixture with a size ratio of 0.3, which is constructed by filling the octahedral holes of an FCC crystal of large hard spheres with small spheres and where the fraction of octahedral holes filled with small spheres can be tuned from 0 to 1 \cite{filion2011self}. In these crystal structures, the small spheres also hop between neighbouring octahedral holes via a tetrahedral hole \cite{filion2011self}. In the HCP structure, the polymers are in a stretched state at low polymer packing fraction and they  display cooperative behavior at high polymer packing fraction, resulting in a bimodal distribution of collapsed and stretched polymer configurations.

It is worthwhile to explore the possibility to experimentally verify the theoretical predictions regarding the stability of the HCP structure over the FCC structure. Our simulation study shows that for the system parameters that we  considered,  the HCP structure is stable with respect to the FCC phase for a polymer reservoir packing fraction $0.225 < \eta_p^r < 0.275$. The  HCP  phase, which is dense in colloids and dilute in polymer coexists with a fluid phase, which is very dilute in colloids and rich in polymers (see Fig. \ref{fig02}b). The packing fraction of the stable HCP crystal structure $\eta_c \geq 0.72$, almost approaches the close-packed density of $\eta_{cp}\simeq 0.74$. Typical experimental samples of colloids are polydisperse in size. As noted by Pusey \cite{Pusey-P.N.1987} for crystals with high packing fractions, the maximum allowed polydispersity $\Delta$ is  constrained by the expression $\Delta\simeq \left(\eta_{cp}/\eta_c \right)^{1/3} - 1$. This empirical relation is based on the fact that the particles with sizes corresponding to  the tails of the size distribution should still fit onto the crystal lattice without distorting the crystal structure. Using this constraint, we find that the size polydispersity $\Delta$ should be less than $1 \%$  in order to achieve a colloidal crystal with a  packing fraction $\eta_c \simeq 0.72$, which  is highly challenging, but not impossible. Another possible complication might be the size of the polymer as given by its radius of gyration that we considered in our model system, which  is approximately on the same order of magnitude as the size of the colloids. Moreover, in a recent series of papers  \cite{mahynski2014stabilizing,mahynski2015relative,mahynski2015tuning} Panagiotopolous and coworkers have investigated how interactions between the polymers and the colloids impact the stability of the HCP structure over the FCC structure. They concluded that in order to stabilize the HCP structure the interactions between the colloid and polymer need  to be hard or repulsive, characterized by a positive second virial coefficient  $B_2$. It is unclear if these two conditions can be realized experimentally. Another recent work of Mahynski et al shows that complex, open (non-closed packed) crystal morphologies can also be stabilized by suitably designing the architecture of the polymer \cite{Mahynski2016}.  In conclusion, it is evident that with a careful fine-tuning of the architecture of the polymer, it is possible to target a particular polymorph among a set of competing structures. We hope that our findings will  inspire new experimental studies in this direction. 

\begin{acknowledgments}
J.R.E. and M.D. acknowledge financial support from a Nederlandse Organisatie voor Wetenschappelijk Onderzoek (NWO) VICI grant.  J.R.E. and M.D acknowledge a NWO-EW grant for computing time in the Dutch supercomputer Cartesius. \end{acknowledgments}

\bibliography{polymers}

\end{document}